\documentclass[conference]{IEEEtran}
\IEEEoverridecommandlockouts

\usepackage{balance}
\usepackage{microtype}
\usepackage{graphicx}
\usepackage{subfigure}
\usepackage{booktabs} 

\usepackage{algorithm}
\usepackage[noend]{algorithmic}
\usepackage{booktabs}
\usepackage{multirow}

\usepackage{booktabs}       
\usepackage{amsfonts}       
\usepackage{nicefrac}       
\usepackage{microtype}      
\usepackage{xcolor}         
\usepackage{graphicx}
\usepackage{subfigure}
\usepackage{amsmath}
\usepackage{amssymb}
\usepackage{mathtools}
\usepackage{amsthm}
\usepackage{paralist}
\usepackage{multirow}
\usepackage{makecell}
\usepackage{wasysym}
\usepackage{caption}
\usepackage{adjustbox}
\usepackage{threeparttable}
\usepackage{lipsum}
\usepackage{cite}
\usepackage{hyperref}

\def\BibTeX{{\rm B\kern-.05em{\sc i\kern-.025em b}\kern-.08em
    T\kern-.1667em\lower.7ex\hbox{E}\kern-.125emX}}
\begin{document}

\title{RobPI: Robust Private Inference \\ against Malicious Client}
\author{\IEEEauthorblockN{Jiaqi Xue}
\IEEEauthorblockA{\textit{University of Central Florida} \\
jiaqi.xue@ucf.edu}
\and
\IEEEauthorblockN{Mengxin Zheng}
\IEEEauthorblockA{\textit{University of Central Florida} \\
mengxin.zheng@ucf.edu}
\and
\IEEEauthorblockN{Qian Lou}
\IEEEauthorblockA{\textit{University of Central Florida} \\
qian.lou@ucf.edu}
}

\maketitle

\begin{abstract}
The increased deployment of machine learning inference in various applications has sparked privacy concerns. In response, private inference (PI) protocols have been created to allow parties to perform inference without revealing their sensitive data. 
Despite recent advances in the efficiency of PI, most current methods assume a semi-honest threat model where the data owner is honest and adheres to the protocol. However, in reality, data owners can have different motivations and act in unpredictable ways, making this assumption unrealistic. To demonstrate how a malicious client can compromise the semi-honest model, we first designed an inference manipulation attack against a range of state-of-the-art private inference protocols. This attack allows a malicious client to modify the model output with 3$\times$ to 8$\times$ fewer queries than current black-box attacks. Motivated by the attacks, we proposed and implemented RobPI, a robust and resilient private inference protocol that withstands malicious clients. RobPI integrates a distinctive cryptographic protocol that bolsters security by weaving encryption-compatible noise into the logits and features of private inference, thereby efficiently warding off malicious-client attacks. Our extensive experiments on various neural networks and datasets show that RobPI achieves $\sim 91.9\%$ attack success rate reduction and increases more than $10\times$ query number required by malicious-client attacks. 
\end{abstract}


\section{Introduction}
Machine-learning-as-a-service (MLaaS) is a powerful method to provide clients with intelligent services and has been widely adopted for real-world applications~\cite{LIU201711}, such as image classification/segmentation for home monitoring systems~\cite{kuna2017,wyze}, intrusion detection~\cite{ASHIKU2021239}, fraud detection~\cite{dl:fraud_detection, Delphi:usenix2020}. Nevertheless, the integration of MLaaS into many such applications engenders privacy concerns~\cite{CryptoNets:ICML2016, Delphi:usenix2020, Lou:ICLR2021:safenet}. For instance, home monitoring systems like Kuna and Wyze utilize neural networks to categorize objects in user home video feeds, such as vehicles stationed near the user's residence or identifying visitors' faces, which may intrude on personal privacy. 

To address these privacy concerns, numerous recent studies, as illustrated in Table~\ref{t:PNet_label}, have put forth protocols for cryptographic prediction, specifically, private inference over (convolutional) neural networks, by leveraging various cryptographic primitives, e.g., fully homomorphic encryption (FHE)~\cite{FHE, xue2024cryptotrain, kumar2025tfhe, zhang2025cipherprune, zheng2023priml, xue2025sok, xuedictpfl}. A private inference protocol facilitates interaction between data owners and model owners, denoted by PNet, in a way that allows the user to receive the prediction outcome while simultaneously guaranteeing that neither party obtains any additional information pertaining to the user's input or the model's weight parameters. Numerous studies, for instance, \cite{Brutzkus:ICML19, FCryptoNets:arxiv19, CryptoDL:arxiv17, CHET, Lou2021HEMETAH, TenSEAL, HeLayers, cryptoeprint}, as indicated in Table~\ref{t:PNet_label}, have achieved these assurances. However, all these studies make an assumption of semi-honest protocol adherence, implying that both the client-side data owner and server-side model owner comply with the protocol without malicious behaviors.

\begin{figure}[t!]
  \centering
    \includegraphics[width=\linewidth]{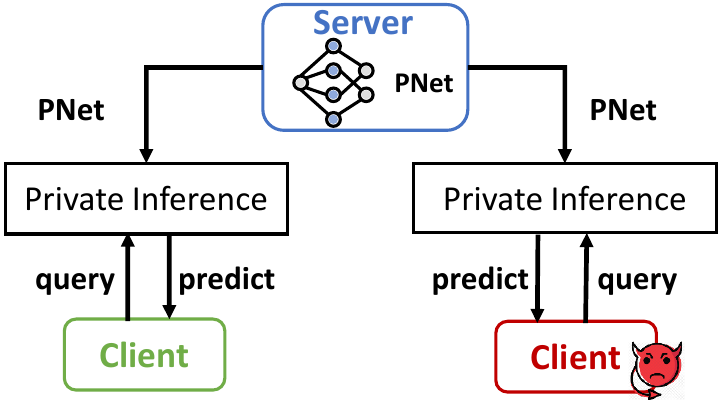}
    \caption{Client may be malicious in private inference.}
    \label{fig:overview}
\end{figure}

\begin{table}[h!]
\small
\begin{tabular}{lcc}
\toprule
& \multirow{2}{*}{\makecell[c]{vulnerable to\\adversarial examples }} & \multirow{2}{*}{\makecell[c]{requires network\\modification}} \\
& & \\
\midrule
LoLa & \CIRCLE & \CIRCLE \\
CryptoNets & \CIRCLE & \CIRCLE \\
CryptoDL & \CIRCLE & \CIRCLE \\
CHET & \CIRCLE & \CIRCLE \\
HEMET & \CIRCLE & \CIRCLE \\
TenSEAL & \CIRCLE & \CIRCLE \\
HeLayers & \CIRCLE & \CIRCLE \\
\textbf{RobPI (ours)} & \Circle & \Circle \\
\bottomrule
\end{tabular}
\begin{tablenotes}
\begin{small}

\item \CIRCLE=\text{provides property}
\item \Circle=\text{does not provide property}
\end{small}
\end{tablenotes}
\caption{Prior private inferences are vulnerable to malicious-client adversarial examples and need network modification for higher robustness.}
\label{t:PNet_label}
\end{table}


While the majority of the literature adopts the semi-honest threat model, it is fundamentally less probable that all clients will adhere to proper behavior. The server is hosted by a single service provider, and existing cloud providers employ strict access control, and physical security measures, making it considerably challenging to circumvent these safeguards. Furthermore, if a service provider is found to be acting maliciously, the repercussions could be severe due to public accountability~\cite{server-protection,cryptoeprint}. In contrast, clients are numerous, operate on diverse setups under user control, have varying motives, and it suffices that one behaves maliciously, as Figure~\ref{fig:overview} shows. The incentives for a client to cheat are substantial: service providers offer high-stakes services such as intrusion detection, home monitoring systems, and fraud detection. Consequently, a client, who may be an attacker, could seek to obtain unauthorized access to information or manipulate the decision-making processes of the MLaaS system to attain personal or financial advantages. The recent study, MUSE~\cite{cryptoeprint}, investigates malicious clients attempting to pilfer models through queries, but it does not offer solutions for adversarial output manipulation attacks.

To highlight the dangers associated with a malicious client, we introduce a new adversarial output manipulation attack against semi-honest private inference protocols, i.e., PI-ATTACK. This attack enables a malicious client to modify the model's output using a reduced number of inference queries. With an efficiency improvement of $3\times \sim 8\times$ compared to the most effective prediction manipulation attacks for plaintext inference~\cite{guo2019simple, attack2}, this attack demonstrates that the utilization of semi-honest private inference protocols can significantly enhance a malicious client's ability to alter model predictions.

To counteract such amplification of security risk, we introduce RobPI, a robust private inference protocol designed to function effectively under a client-malicious threat model. In this context, the server is presumed to maintain semi-honest behavior, while the client may diverge substantially from the protocol's stipulations. As we will elaborate, adopting this model empowers RobPI to surpass current state-of-the-art methods in its performance efficiency. 

\noindent\textbf{Contributions.} We summarize our contributions as follows:
(i) We show an inference manipulation attack, denoted by \textit{PI-Attack}, against private inference protocols.  This attack enables a malicious client to manipulate the model's output with few queries.  (ii) We introduce \textit{RobPI}, a robust private inference protocol resilient to malicious clients. In designing \textit{RobPI},  we propose to add cryptography-compatible noise in the features and logits layer. In addition, we introduce a dynamic noise training (DNT) technique to further improve the resilience against malicious clients. \textit{RobPI} increases more than $10\times$ query numbers compared to prior defense methods, which increases the attack difficulties. (iii) We provide theoretical analysis on \textit{RobPI} and our implementation of \textit{RobPI} is able to decrease attack success rate by $\sim 91.88\%$ against malicious clients on various neural networks and datasets.

\section{Background and Related Works}

\noindent\textbf{Private Inference.}
One popular private inference paradigm is based on fully homomorphic encryption (FHE)~\cite{FHE}, which stands out as FHE-based private inference allows non-interactive neural network inference privacy-preserving, i.e., inference on encrypted data without needing to decrypt it~\cite{lou2019she, lou2020falcon, lou2020autoprivacy, lou2021efficient, lou2025secure, zheng2022cryptolight, kumar2026fhaim,xue2025sok}. Compared to interactive private inference based on multi-party computation~\cite{Delphi:usenix2020,zheng2023primer, zhang2024heprune,zhang2025cipherprune, feng:2020:cryptogru,zhang2026cryptogen}, FHE-based private inference has two main advantages: (1) non-interactive end-to-end inference and (2) more secure against layer-by-layer attacks, e.g., client-malicious model extraction attacks. In particular, non-interactive inference is more friendly to clients who have no powerful machines or high-bandwidth network connections. This is because interactive private inference has the drawback that the clients must stay online during the computation. We noticed that many protocols of private inference in Table~\ref{t:PNet_label}, such as \cite{HeLayers, cryptoeprint}, share a similar workflow where the client sends encrypted data to a server. The server then transforms a regular inference (RI) into a Privacy-preserving Inference (PI), which allows inference on encrypted data without decryption. The original real-number convolution in the RI is replaced with a fixed-point FHE convolution, and the non-linear $ReLU$ function is swapped for polynomial-approximated functions like the $square$ function. The result remains encrypted and can only be decrypted by the data owner with a private key, ensuring privacy preservation.  

\noindent \textbf{Black-box Inference Manipulation Attacks.} 
The malicious-client attack operates on the principle that the attacker can access the black-box MLaaS based on private inference, thereby gaining the capability to manipulate input data to achieve unauthorized access or influence decisions derived from private inference. Current black-box inference manipulation attacks~\cite{guo2019simple, attack2}, have been validated as effective in producing adversarial examples that can control the inference output without the necessity for retraining a surrogate model. Specifically, these methods illustrate how a mathematical tool - the discrete cosine transform (DCT) - can be employed to shift an image from the spatial domain to the frequency domain. By initiating a search from lower frequencies and progressing to higher ones, one can effectively pinpoint an adversarial sample, thereby reducing the number of required queries. Our \textit{PI-Attack} strategy follows this DCT-based method~\cite{guo2019simple, attack2}.

\noindent\textbf{Resilient Neural Networks.}
To defend against query-based black-box attacks, 
\cite{salman2020denoised, byun2021small} show that adding random noise into the input~\cite{qin2021random} or model~\cite{byun2021small} can defend against attacks without perceiving the inputs. Also, R\&P~\cite{xie2017mitigating} proposes an input random-transform defense method. RSE~\cite{liu2018towards} adds large Gaussian noise into both input and activation and uses ensembles to avoid an accuracy decrease. PNI~\cite{he2019parametric, cohen2019certified, salman2019provably} incorporates noise in the training. However, these defense methods sacrifice enormous accuracy. And the input-transform function in R\&P and the ensemble method in RSE introduce a large overhead for PI. \cite{rusak2020simple} introduces that the model with Gaussian augmentation training could defend the common corruptions. RND~\cite{qin2021random} extends the methods in~\cite{rusak2020simple,byun2021small} and achieves the state-of-the-art defense against black-box attacks. However, RND does not consider the distinct features of PI, i.e., quantized activation and model, and polynomial activation that has a significant decay on the added noise of the input, thus restricting the defense effect on private inference.

\begin{figure}[h!]
\centering
\includegraphics[width=\linewidth]{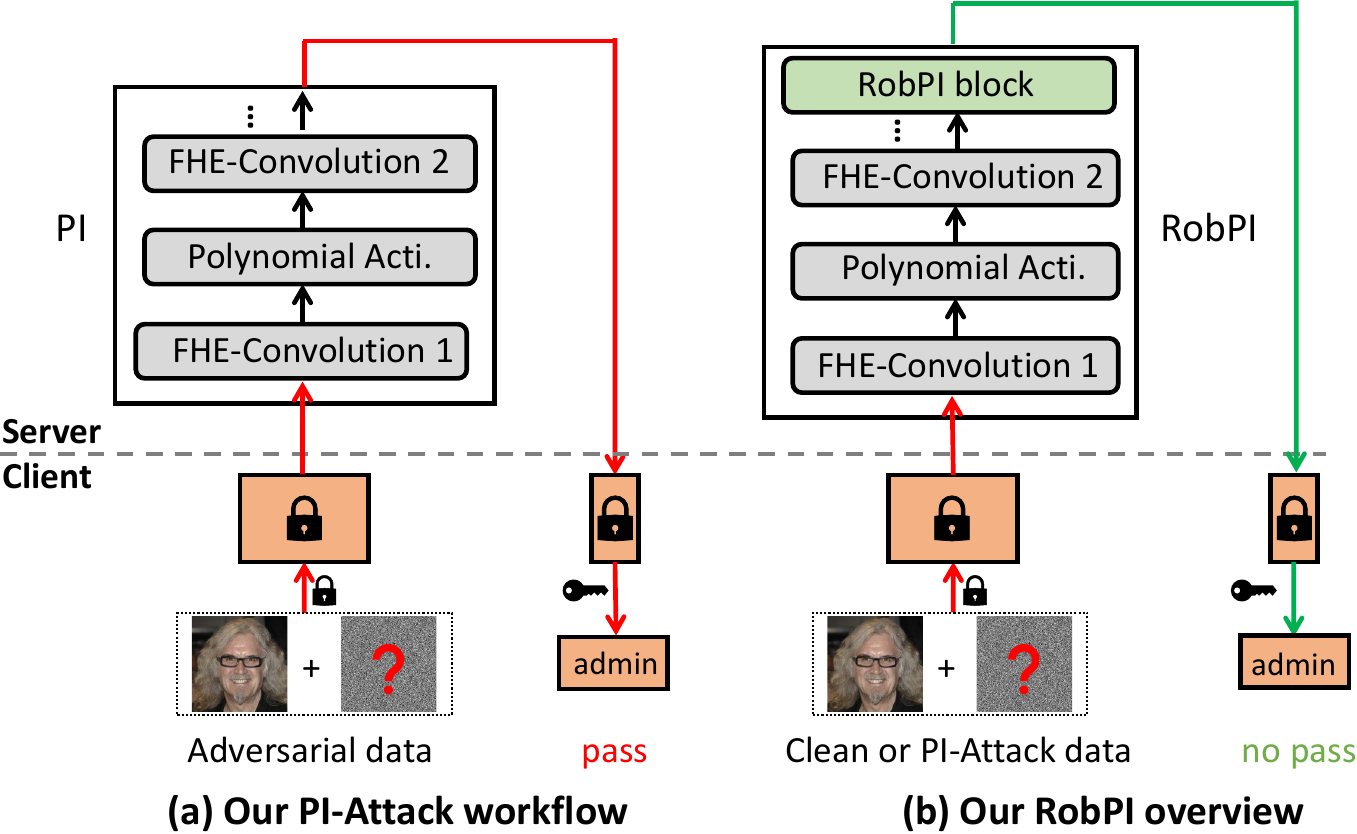}
\caption{(a) PI-Attack use case on the privacy-preserving face recognition system. (b) Our RobPI enables a fast, accurate, and robust PI. }
\label{fig:motivation-ICLR}
\end{figure}
\section{PI-Attack: Attacks on Private Inference}
\label{s:PI-Attack}

\noindent\textbf{Attack Threat Model.} Our attack strategies are designed to manipulate the output of private inference to gain unauthorized access, potentially resulting in personal or financial benefits, as seen in intrusion detection systems~\cite{wyze,kuna2017} and fraud detection systems~\cite{dl:fraud_detection}. These attacks target semi-honest private inference protocols that possess a specific characteristic: the client's final output should coincide with the plaintext output of the final linear layer. Several private inferences, including all the current FHE-based methods~\cite{Brutzkus:ICML19, FCryptoNets:arxiv19, CryptoDL:arxiv17, CHET, Lou2021HEMETAH, TenSEAL, HeLayers, cryptoeprint}, satisfy this condition. 
In Figure~\ref{fig:motivation-ICLR}(a), we take a privacy-preserving face-recognition authentication system as an example to illustrate more about our threat model use case, where a non-admin client attacker aims to obtain admin permission. To achieve this goal, the attacker has a motivation to find an adversarial noise or example that causes the system to produce a target-class prediction, e.g., admin, in order to bypass the recognition system.


\begin{algorithm}[t!]\small
    \caption{PI-Attack in Pseudocode}
    \label{alg:attack} 
    \begin{algorithmic}[1]
        \STATE $\textbf{Input:}$ image $x \in \mathbb{R}^{d\times d \times c}$, label $y$, step size $\epsilon$.
        \STATE adversarial perturbation $\delta = 0$
        \STATE $\mathbf{O} = M_p(x)$, $t=0$
        \STATE $\hat{x} = DCT(x)$   \;\;   \# for each channel
        \WHILE{$ \mathbf{O}_y = max_{y'}\mathbf{O}_{y'} \; and \;t< d^2 $}
        \STATE get $\hat{x_{i,j}}$ with the lowest frequency from  $\hat{x}$. 
        \STATE $\hat{x}=\hat{x}.pop(\hat{x_{i,j}})$ 
        \STATE$Q = Basis(\hat{x_{i,j}})$
         \FOR{$\hat{\alpha_t} \in \{\lambda_t \cdot \epsilon, -\lambda_t\cdot \epsilon \} $} 
          \STATE $t++$
          \STATE $\mathbf{O'} =  M_p(x+\delta_t + IDCT(\hat{\alpha_t} \cdot Q)) $
        \IF{$sign(O^{'}_{y}-O_y)<0$} 
        \STATE $\delta_{t+1} = \delta_t + IDCT(\hat{\alpha_t} \cdot Q)$
        \STATE $\mathbf{O} = \mathbf{O'} $
        \STATE {$\textbf{break}$}
        \ENDIF
        \ENDFOR
        \ENDWHILE	
    \STATE $\textbf{Return }${$\delta$}
    \end{algorithmic}
    
\end{algorithm}

\noindent\textbf{Attack Strategy.}
To efficiently manipulate the inference, we propose a PI-Attack method that is optimized for PNet shown in Algorithm~\ref{alg:attack}. The PI-Attack method takes one clean image $X\in \mathbb{R}^{d\times d \times c}$, true label $y$, and step size $\epsilon$ as inputs, and generates adversarial perturbation $\delta$, where $d$ is the input width or height, $c$ is channel number. We define the prediction score probability of PNet model as $\mathbf{O}=M_p(x)$. Instead of adding perturbation in the spatial domain, we adopt a more efficient search direction $Q$ in the frequency domain by the existing DCT tool and convert the frequency domain perturbation $\hat{\alpha_t} \cdot Q$ back to the spatial domain by inverse DCT (IDCT). DCT and IDCT are defined in the Appendix. The key idea of the algorithm is simple, i.e., for any direction $Q$ and step size $\hat{\alpha_t}$, one of $x+IDCT(\lambda_t \cdot \epsilon \cdot Q)$ or $x+IDCT(-\lambda_t \cdot \epsilon \cdot Q)$ may decrease $\mathbf{O}=M_p(x)$. We iteratively pick direction basis $Q$ in the ascending order of frequency value $\hat{x_{i,j}}$ in $\hat{x}$. Note that we randomly sample one $\hat{x_{i,j}}$ when there are multiple entries with the same value. For each query $t$, if the prediction probability $\mathbf{O'}$ of $x+IDCT(\lambda_t \cdot \epsilon \cdot Q)$ is decreased over $\mathbf{O}$, we will accumulate the perturbation $\delta_{t}$ with it, otherwise, we will accumulate the $IDCT(-\lambda_t \cdot \epsilon \cdot Q)$.

The search efficiency of PI-Attack algorithm is mainly dependent on two components, i.e., arc-shaped search order $Q$ in the frequency domain and perturbation size schedule $\lambda_t$.  In particular, frequency domain input $\hat{x}$ is calculated by $DCT(x)$ for each channel, where the top-left positions of $\hat{x}$ have lower frequency values. Since low-frequency subspace adversarial directions have a much higher density than high-frequency directions, we try to perform the search from lower frequency to higher frequency before a successful attack. To achieve this goal, we iteratively extract the value $\hat{x_{i,j}}$ with the lowest frequency from $\hat{x}$. To avoid the repeating search, we pop out the $\hat{x_{i,j}}$ from the remaining search space $\hat{x}$ by $\hat{x} = \hat{x}.pop(x)$ shown in Algorithm~\ref{alg:attack}. The search direction basis $Q$ is set as $\hat{x_{i,j}}$ for the $t$-th query, which means that we only add the perturbation in the position of $\hat{x_{i,j}}$ and check if it decreases the prediction probability at the $t$-th query.  
\begin{equation} \small
    \lambda_t=  \lambda_{min} +\frac{1}{2}(\lambda_{max} - \lambda_{min})(1+cos(\frac{t}{T})\cdot \pi)
    \label{eq:schedule}
\end{equation}

Since $\hat{x_{i,j}}$ with lower frequency contains more dense information than high-frequency values, we propose a perturbation size schedule $\lambda_t$ to assign a larger perturbation size to the positions with lower frequency, which further improves the search efficiency. For $t$-th query, the perturbation size $\alpha_t$ is defined as the multiplication between $\lambda_t$ and frequency domain perturbation seed $\epsilon$. We define the cosine annealing schedule $\lambda_t$ in Equation~\ref{eq:schedule}, where $\lambda_{min}$ and $\lambda_{max}$ are the minimum and maximum coefficients of perturbation size, respectively, and $\lambda_t\in [\lambda_{min}, \lambda_{max}]$, $T$ is the query range cycle.

\begin{figure}[h!]
  \centering
   \includegraphics[width=0.9\linewidth]{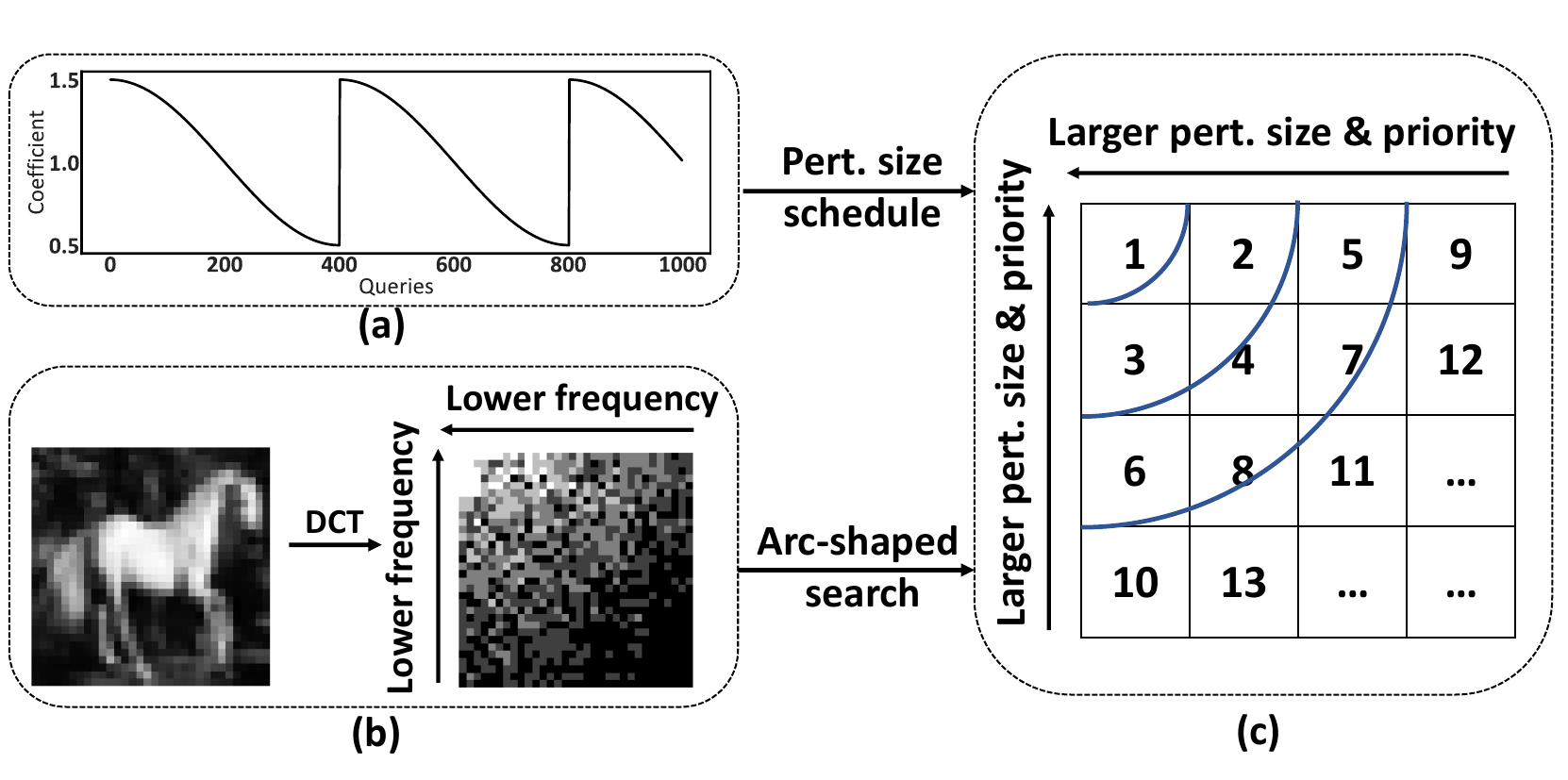}
   \caption{PI-Attack workflow.}
   \label{fig:attack_methods}
\end{figure}

In Figure~\ref{fig:attack_methods}, we illustrate the proposed PI-Attack workflow. In particular, Figure~\ref{fig:attack_methods}(b) describes the conversion from input $x$ into frequency domain $\hat{x}$ whose values decide the search priority in our Algorithm~\ref{alg:attack}, i.e., PI-Attack. For example, the top-left position is of the highest priority to add perturbation, thus we perform the perturbation search on it in the first query. In Figure~\ref{fig:attack_methods}(c), we list the number of search orders, i.e., query orders. For instance, in the second query, we add a perturbation on the element in the first row, second column. This search order forms an arc shape, so we call this search order an arc-shaped search. Figure~\ref{fig:attack_methods}(a) demonstrates the perturbation schedule, i.e., the schedule size coefficient $\lambda_t$ over the query order. For example, the first query with the lowest frequency uses the largest $\lambda_t$ and the following queries with larger frequency will use smaller $\lambda_t$.

\section{RobPI: Robust Private Inference against Malicious Client}
\label{s:RobPI}

\noindent\textbf{Defense Model.} In the context of protocol adherence, a semi-honest adversary adheres strictly to the prescribed protocol, yet attempts to extract information about other parties' inputs by analyzing the messages it receives. On the other hand, a malicious adversary is not bound by these rules and has the freedom to deviate from the protocol in any way it sees fit. We design RobPI to defend the new threat model named \textit{security against malicious clients}. We use Figure~\ref{fig:motivation-ICLR}(b) to show our RobPI will defend against this attack, so that the attacker can no longer fool the recognition system.

\noindent\textbf{RobPI Design Principle.} In contrast to NN used for plaintext inference, the private inference-designed PNet exhibits unique characteristics, such as polynomial activation and quantized parameters. Consequently, directly adapting previous work~\cite{qin2021random}—which injects Gaussian noise into the input layer—may not be optimal for PNet. We noticed that previous work~\cite{qin2021random} displays a diminished defense impact when applied to private inference. This can be attributed to the significant attenuation of the injected noise by the multi-layer polynomial-approximated activation in PNet, such as the $square$ function, rendering the defense noise virtually ineffective on the model output, especially for deeper neural networks. The diminished noise cannot significantly influence the final-layer logits, thus reducing the defense performance of prior work on PNet. In \textit{RobPI}, we propose a simple, efficient, and provably effective method that circumvents the noise decay issue by adding noise to the output layer. Nonetheless, this strategy might be susceptible to an \textit{average inference attack}—a phenomenon we detail in Appendix—particularly when the injected noise has a zero mean. To counteract this, one can simply use non-zero-mean noise and integrate it into the final two layers of the network. To further augment the effectiveness of \textit{RobPI}, we also introduce a novel dynamic noise training (DNT) technique.

In query-based inference manipulation attacks, the aggressor iteratively introduces a minor disruption to the input, subsequently inspecting whether consecutive queries yield varying prediction probabilities. If the probability of the objective prediction for the ($t+1$)-th query diminishes compared to the $t$-th query, the introduced perturbation is retained. Conversely, the attacker removes the adversarial disruption. The efficiency of the search heavily relies on the correct determination of the perturbation search direction, i.e., whether to add or subtract the disruption in each query. Consequently, a defender can reduce the efficiency of the attack by perturbing the perturbation search direction, thus misleading the prediction probabilities. Inspired by these observations, we propose a swift and accurate defense methodology that simply introduces noise to the output probability in each query, resulting in a robust RobPI. Our RobPI defense approach is designed to achieve two objectives: (i) ensuring that the introduced defense noise doesn't significantly alter the prediction probability of the normal dataset, thus maintaining accuracy, and (ii) ensuring that the introduced defense noise significantly disrupts the attack search direction, thereby reducing the search efficiency. 

\noindent\textbf{RobPI Defense Formulation.}
We use Equation~\ref{eq:Apx} to define the prediction prabability difference of two queries on PNet, $M_p(x+\delta_{t}+\mu_t)$ and $M_p(x+\delta_t)$, where $\delta_t$ is the accumulated perturbation at $t$-th query, $\mu_t$ is the perturbation of $t$-th query, e.g., $IDCT(\hat{\alpha_t} \cdot Q)$ if defending PI-Attack in Algorithm~\ref{alg:attack}. 
\begin{equation} \small
     \label{eq:Apx}
    A_p(x,t)=M_p(x+\delta_t+\mu_t)- M_p(x+\delta_t)
\end{equation}
In Equation~\ref{eq:Dpx}, we define the main step of the proposed RobPI defense method. One can see that two noises $\sigma \Delta_{t+1}$ and $\sigma \Delta_{t}$ are added into $(t+1)$-th query and $t$-th query, respectively, to disturb the attack search direction. Those noises are sampled from the same standard Gaussian distribution $\Delta \sim \mathcal{N}(0,1)$ and multiplied by a small factor $\sigma$. Note that the added noise shares the same encoding method with PNet for correct decryption of prediction result. The key idea of adding noise in the query result is to disturb the difference, i.e., $A_p(x,t)$, of two attack queries and mislead the search directions.
\begin{equation} \small
\label{eq:Dpx}
\begin{aligned}
        D_p(x,t) &= (M_p(x+\delta_{t}+\mu_t)+ \sigma \Delta_{t+1})- (M_p(x+\delta_t) +\sigma \Delta_t) \\
        &=A_p(x,t)+\sigma (\Delta_{t+1}-\Delta _t)
\end{aligned}
\end{equation}
Specifically, the disturbance success happens when the signs of $A_p(x,t)$ and $D_p(x,t)$ are different. We use Equation~\ref{eq:Sx} to define the probability of disturbance success (DSP). A higher DSP will induce a lower attack success rate (ASR). Therefore, it is of great importance to understand the factors impacting the DSP. 


\begin{equation} \footnotesize
\label{eq:Sx}
    S(x,t)= P(sign (A_p(x,t)) \neq sign(D_p(x,t)))
\end{equation}

\begin{figure}[h!]
  \centering
   \includegraphics[width=\linewidth]{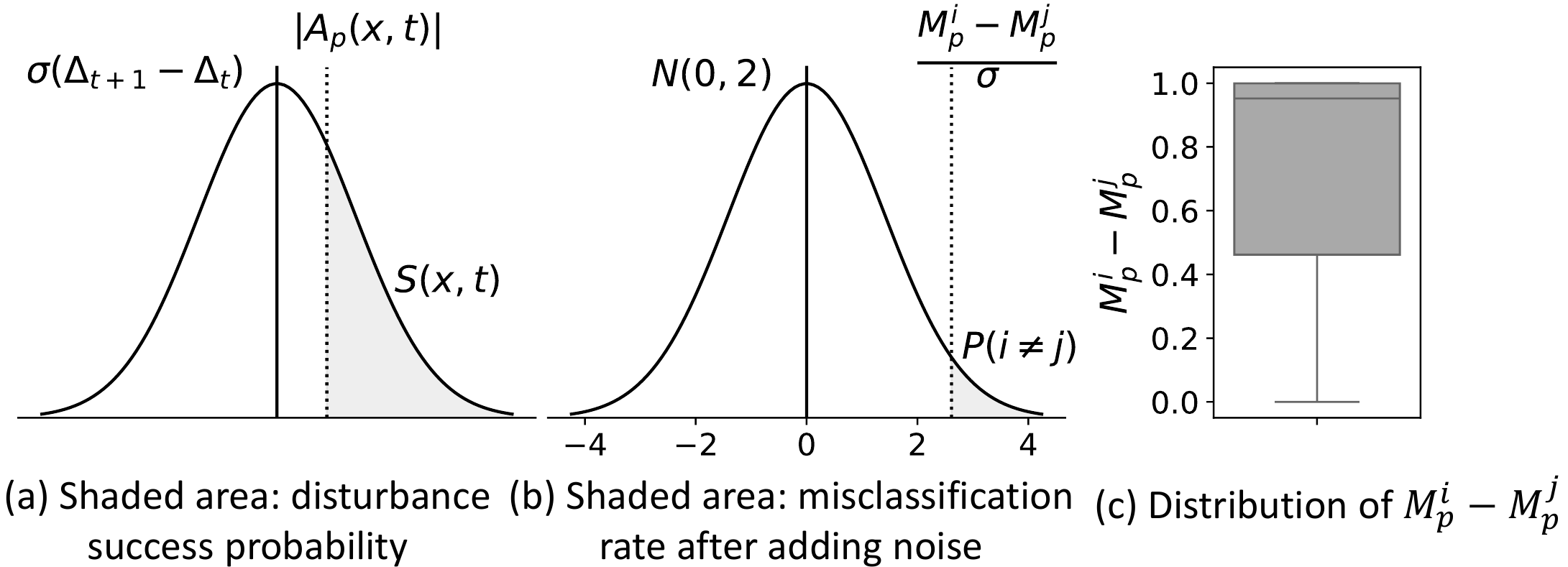}
   \caption{\textcolor{black}{(a) Disturbance success probability S(x, t) for one query. (b) incorrect prediction rate $P(i\neq j)$, which means the probability of misclassification after adding noise on the confidence scores. (c) distribution of the difference between the highest predicted score $M_p^i$ and the second highest predicted score $M_p^j$.}}
   \label{fig:math}
\end{figure}

We theoretically analyze and calculate the DSP in Equation~\ref{eq:SxValue}. According to Equation~\ref{eq:Dpx}, the only difference of $D_p(x,t)$ and $A_p(x,t)$ is $\sigma (\Delta_{t+1}-\Delta _t)$, thus $S(x,t)$ is equal to the probability of adding $\sigma (\Delta_{t+1}-\Delta _t)$ to change the sign of $A_p(x,t)$. Given the Gaussian distribution $\sigma(\Delta_{t+1}-\Delta _t)\sim \mathcal{N}(0,2\sigma^2)$, the $S(x,t)$ is equal to $1- \phi (|A_p(x,t)|;0, 2\sigma^2)$, where $\phi ()$ is the cumulative distribution function (CDF) of Gaussian distribution. This is because if $A_p(x,t)<0$, the added noise sampled from $\mathcal{N}(0,2\sigma^2)$ should be larger than $|A_p(x,t)|$ to change the sign of $A_p(x,t)$, thus its probability is $1- \phi (|A_p(x,t)|;0, 2\sigma^2)$; otherwise, the added noise should larger than $|A_p(x,t)|$ to change the sign of $A_p(x,t)$, thus the probability is also $1- \phi (|A_p(x,t)|;0, 2\sigma^2)$. Therefore, using the CDF equation, one can calculate the DSP in Equation~\ref{eq:SxValue}, where $erf$ is Gauss error function. We demonstrate in Equation~\ref{eq:SxValue} that DSP is impacted by two factors, i.e., $|A_p(x,t)|$ and $\sigma$. DSP has a positive relationship with $\sigma$ but is negatively relative to $|A_p(x,t)|$. In Figure~\ref{fig:math} (a), we use the shaded area to illustrate the value of $S(x,t)$.
\begin{equation} \small
\label{eq:SxValue}
\begin{aligned}
    S(x,t) &= 1- \phi (|A_p(x,t)|;\mu, 2\sigma^2 )\\
    &= 1-\frac{1}{2\sigma \sqrt{\pi}} \int_{-\infty}^{|A_p(x,t)|}exp(\frac{(|A_p(x,t)|-\mu)^2}{-4\sigma^2})d|A_p(x,t)| \\
    &=\frac{1}{2}-\frac{1}{2}erf(\frac{|A_p(x,t)|}{2\sigma}) \quad \text{where} \quad \mu=0
\end{aligned}
\end{equation}

\noindent\textbf{RobPI Analysis.}
We theoretically analyze the effects of our RobPI defense method on clean accuracy. When applying our PNet on a $n$-class classification task, we can define prediction score as $\mathbf{O} = \{M_p^0, M_p^1, ..., M_p^{n-1}\}$ for clean data. Since our defense method adds Gaussian noise $\sigma \Delta_t$ to the $\mathbf{O}$, we define the prediction score after our defense method as $\mathbf{O^{\sigma}} = \{M_p^0+\sigma \Delta_t^0, M_p^1+\sigma \Delta_t^1, ..., M_p^{n-1}+\sigma \Delta_t^{n-1}\}$. The classification results of $\mathbf{O}$ and $\mathbf{O^{\sigma}}$ are $i=argmax(O)$ and $j=argmax(O^{\sigma})$, respectively. Therefore, RobPI will predict an incorrect classification if $i\neq j$. We use Equation~\ref{eq:Pij} to describe the probability of $P(i \neq j)$ that is positively relative to $\sigma$ but negatively relative to $M_p^i -M_p^j$. In Figure~\ref{fig:math} (b), we use the shaded area to illustrate the value of $P(i\neq j)$. Our RobPI achieves a tiny $P(i\neq j)$ and a large $S(x,t)$ given a small $\sigma$, therefore obtaining an accurate and robust PNet. 
Figure~\ref{fig:math} (c) demonstrates the distribution of $M_p^i -M_p^j$ and most of the values are larger than $0.5$ on CIFAR-10. The $\frac{M_p^i -M_p^j}{\sigma} >5$ since the $\sigma$ value is $<0.1$. Those observations show that $P(i\neq j)$ is tiny since 1-$\phi(\frac{M_p^i -M_p^j}{\sigma} >5;0,\sigma^2=2)$ is near zero. Therefore, the defender can adjust $\sigma$ based on the validation accuracy and misclassification budget obtained in the training to achieve better defense effectiveness according to Equation \ref{eq:Pij}.

\begin{equation} \small
\label{eq:Pij}
\begin{aligned}
    P(i\neq j) &= P ((M_p^i + \sigma\Delta_t^i)< (M_p^j + \sigma\Delta_t^j))\\
    &= P(\frac{M_p^i -M_p^j}{\sigma}<\Delta_t^j-\Delta_t^i)\\
    &=\frac{1}{2}-\frac{1}{2}erf(\frac{M_p^i -M_p^j}{2\sigma})
\end{aligned}
\end{equation}


\noindent\textbf{RobPI with dynamic noise training (DNT).} By analyzing the Disturbance success probability $S(x,t)$ in Equation~\ref{eq:SxValue} and clean accuracy decrease rate $P(i\neq j)$ after applying our RobPI defense method, we reveal that a larger $\sigma$ will improve the defense effect but also may decrease the clean accuracy. To further avoid the clean accuracy decrease, one can reduce the noise sensitivity of PNet model or enlarge the difference between $M_P^i$ and $M_P^j$ in Equation~\ref{eq:Pij}. Inspired by those observations, we additionally equip our RobPI with dynamic noise training, denoted by RobPI-DNT, to enable a better balance between clean accuracy and defense effects. We use Algorithm~\ref{alg:RobPI-DNT} to describe RobPI-DNT that adds dynamic epoch-wise Gaussian noise $\sigma_i\Delta_t$ during each training iteration. Our results in Table~\ref{t:results_defense} show RobPI-DNT attains higher clean accuracy over RobPI. 

\begin{algorithm}[h!]\small
\caption{RobPI with DNT}
\label{alg:RobPI-DNT} 
    \begin{algorithmic}[1]
        \STATE $\textbf{Input:}$ RobPI model $M_p$, training data ($x, y$).
        \STATE  $t= 0$
        \FOR{$i=1$ to $epochs$}
        \STATE Randomly sample $\sigma_i \in [0, \sigma_{max}]$
        \FOR{$j=1$ to $iterations$} 
        \STATE $pred \; = M_p(x+\sigma_i \Delta_t)$
        \STATE $t++$
         \STATE minimize $loss(pred,y)$ 
       \STATE  update  $M_p$
          \ENDFOR
        \ENDFOR
    \STATE $\textbf{Return }$ {$M_p$}
    \end{algorithmic}
    
\end{algorithm}

\begin{table*}[ht!]
    \centering
\footnotesize
\setlength{\tabcolsep}{5pt}
\caption{The attack comparisons of PI-Attack and prior works, e.g, SimBA-DCT~\cite{guo2019simple} and Square attack~\cite{attack2} on CIFAR-10 and medical dataset.
Untar., Target means untarget and target attacks, respectively.
}
\begin{tabular}{lcccccccccccc}\toprule
\multirow{3}{*}{Schemes} & \multicolumn{6}{c}{CIFAR-10} & \multicolumn{6}{c}{Diabetic Retinopathy} \\
\cmidrule(lr){2-7}\cmidrule(lr){8-13}
 & \multicolumn{2}{c}{Average Queries} & \multicolumn{2}{c}{Average $\ell_2$} & \multicolumn{2}{c}{Success Rate} & \multicolumn{2}{c}{Average Queries} & \multicolumn{2}{c}{Average $\ell_2$} & \multicolumn{2}{c}{Success Rate} \\
\cmidrule(lr){2-3}\cmidrule(lr){4-5}\cmidrule(lr){6-7}\cmidrule(lr){8-9}\cmidrule(lr){10-11}\cmidrule(lr){12-13}
 & Untar. & Target & Untar. & Target & Untar. & Target & Untar. & Target & Untar. & Target & Untar. & Target \\
\midrule
Square &$100.1$ & $301.6$ & $4.21$ & $5.44$ & $85.64\%$ & $71.56\%$ &$50.0$ & $99.1$ & $1.47$ & $1.18$ & $64.32\%$ & $64.14\%$ \\
SimBA-DCT & $101.6$ & $302.4$ &$2.86$ & $4.81$ & $78.28\%$ & $73.98\%$ & $51.8$ & $101.0$ & $0.82$ &$0.92$ & $73.28\%$ & $51.49\%$ \\
PI-Attack & $103.4$ & $299.4$ & $2.79$ & $4.09$ & $81.33\%$ & $81.48\%$ &$50.5$ & $102.5$ & $0.84$ &$0.87$ & $84.36\%$ & $63.28\%$ \\
+Schedule & $99.8$ & $201.5$ & $3.61$ & $4.87$ & {$\mathbf{94.38}\%$} & {$\mathbf{94.22}\%$} & $50.4$ & $98.5$ & $1.15$ & $1.31$ & $\mathbf{89.92}\%$ & $\mathbf{76.48}\%$ \\
\bottomrule
\end{tabular}
\label{t:results_attack}
\end{table*}
\section{Experimental Setup}

\noindent\textbf{Datasets and Models.}
Aligned with recent non-interactive private inference studies \cite{Lou2021HEMETAH, CHET, HeLayers}, we perform our experiments on MNIST, CIFAR-10, and Diabetic Retinopathy~\cite{medical_dataset}. For MNIST, we adopt a network with a convolution block and two fully connected layers, as described in HeLayers~\cite{HeLayers}. For other datasets, we use a structure with three convolution blocks and two fully connected layers. Networks are quantized into 8 bits for MNIST, 10 bits for CIFAR-10, and 16 bits for medical datasets.

\noindent\textbf{Evaluation Metrics.}
Attack Success Rate (ASR): This is the proportion of successful attacks out of the total evaluated images. Higher ASR signifies better attack performance.
Average Queries: This represents the mean number of queries for each evaluated image, calculated by dividing the total number of queries by the total number of images. Fewer average queries suggest a more efficient attack.
Average $\ell_2$ Norm: This is the mean $\ell_2$ norm for each adversarial image, derived by dividing the total $\ell_2$ norm by the total number of evaluated images. A lower average $\ell_2$ norm indicates a smaller adversarial perturbation.
Defense Success Rate (DSR): This is the proportion of successful defenses, equivalent to the attack failure rate. A higher DSR signifies superior defense effectiveness.
Clean Accuracy (ACC): This measures the model's accuracy on the non-adversarial (clean) data.
Disturbance Success Probability (DSP): This is the likelihood of successfully disrupting the attack search direction. 

\noindent\textbf{Methodologies Study.} For attack methods, we compare prior works including SimBA-DCT~\cite{guo2019simple} and Square attack~\cite{attack2} with our PNet-Attack without a schedule and with a schedule in Figure~\ref{fig:attack} and Table~\ref{t:results_attack}. For defense methods, we compare prior works RND, its variant RND-GF~\cite{qin2021random} and Adversarial Training~\cite{goodfellow2014explaining} with our techniques including RobPI, RobPI with input noise, and RobPI-DNT. RobPI simply adds noise in the output layer shown in Equation~\ref{eq:Dpx}. RobPI+Input noise means adding the noise with a different scaling factor in the input layer. RobPI-DNT further incorporates the DNT technique.

\noindent\textbf{Parameter Settings.}
 We set the maximum number of queries for a single evaluated image as 300/100 for the targeted/untargeted attacks, respectively. For Adversarial Training(AT), we add adversarial examples which is generated by SimBA-DCT, and the ration of adversarial examples is $10\%$ of the entire training data. For PNet-Attack, the cycle of schedule $T$ is 400, $\epsilon$ is 1, and $\lambda_{min}$ is 0.5, $\lambda_{max}$ is 1.5. For RobPI, we set $\sigma$ as 0.1. The scaling factor of input noise is set as $0.05$. For the defense method, the $\sigma_{max}$ is set as 0.25. More experimental settings are included in Appendix. The results of the MNIST are shown in Appendix.


\begin{table*}[ht!]
    \centering
\footnotesize
\setlength{\tabcolsep}{5pt}
\caption{The defense comparisons of RobPI and prior works, e.g, RND~\cite{qin2021random}, RND-GF~\cite{qin2021random} and Adversarial Training(AT)~\cite{goodfellow2014explaining}, on CIFAR-10 and medical dataset.
'+Input noise', and '+DNT' represent adding Gaussian noise into the input layer and using an additional DNT method, respectively, on RobPI.}
\begin{tabular}{lcccccccccccc}\toprule
\multirow{3}{*}{Schemes} & \multicolumn{6}{c}{CIFAR-10} &  \multicolumn{6}{c}{Diabetic Retinopathy} \\
\cmidrule(lr){2-7}\cmidrule(lr){8-13}

 & \multicolumn{2}{c}{Clean Accuracy} & \multicolumn{2}{c}{Average queries} & \multicolumn{2}{c}{Success Rate} & \multicolumn{2}{c}{Clean Accuracy} & \multicolumn{2}{c}{Average queries} & \multicolumn{2}{c}{Success Rate} \\
   \cmidrule(lr){2-3}\cmidrule(lr){4-5}\cmidrule(lr){6-7}\cmidrule(lr){8-9}\cmidrule(lr){10-11}\cmidrule(lr){12-13}
 
 & Untar.  &  Target     &   Untar.    & Target   &  Untar. & Target   & Untar.  &  Target                        &   Untar.  & Target      &  Untar. & Target \\

\midrule
RND & \multicolumn{2}{c}{$72.86\%$} & $199.8$ & $301.1$ & $2.03\%$ & $39.22\%$ & \multicolumn{2}{c}{$66.81\%$} & $48.5$ &$53.1$ & $15.00\%$ & $46.40\%$\\
RND-GF & \multicolumn{2}{c}{$73.71\%$} & $204.2$ & $300.4$ & $10.39\%$ & $56.33\%$ & \multicolumn{2}{c}{$67.73\%$} & $50.2$ & $49.3$ & $10.08\%$ & $59.60\%$\\
AT & \multicolumn{2}{c}{$67.88\%$} & $199.5$ & $302.5$ & $49.08\%$ & $86.20\%$ & \multicolumn{2}{c}{$61.37\%$} & $50.0$ & $51.3$ & $37.36\%$ & $79.05\%$\\
RobPI & \multicolumn{2}{c}{$74.10\%$} & $198.2$ & $299.1$ & $56.17\%$ & $88.04\%$ & \multicolumn{2}{c}{$67.91\%$} & $51.7$ & $50.1$ & $49.77\%$ & $74.14\%$ \\
+Input noise & \multicolumn{2}{c}{$73.53\%$} & $202.7$ & $300.1$ & $49.69\%$ & $83.28\%$ & \multicolumn{2}{c}{$65.82\%$} & $49.3$ & $50.7$ & $36.17\%$ & $74.53\%$ \\
+DNT & \multicolumn{2}{c}{$\mathbf{74.55}\%$} & $199.4$ & $299.7$ & $\mathbf{63.36}\%$ & $\mathbf{91.88}\%$  & \multicolumn{2}{c}{$68.09\%$} & $48.9$ & $52.0$ & $\mathbf{66.41}\%$ & $\mathbf{88.67}\%$ \\
\bottomrule
\end{tabular}
\label{t:results_defense}
\end{table*}

\section{Experimental Results}

\noindent\textbf{PI-Attack Evaluation.}
In Table~\ref{t:results_attack}, the performance of our proposed PI-Attack is compared against that of SimBA-DCT and the Square attack on CIFAR-10 and a medical dataset. For CIFAR-10, the Square attack achieved a targeted attack success rate (ASR) of 71.56\%, requiring an average of 301.6 queries and an average $\ell_2$ norm of 5.44. SimBA-DCT, on the other hand, achieved a slightly higher ASR of 73.98\% with a smaller adversarial size, as indicated by a lower $\ell_2$ norm of 4.81. In comparison, our PI-Attack, even without a perturbation size schedule, improved the ASR by 7.5\% while reducing the average $\ell_2$ perturbation norm by 0.72 compared to SimBA-DCT. The introduction of a perturbation size schedule further improved the ASR of PI-Attack by 12.74\%, achieving an average perturbation $\ell_2$ norm of 4.87 with only 201.5 average queries. This is a 22.66\% ASR improvement, a 0.61 reduction in average $\ell_2$ norm, and a reduction of 100.1 queries compared to the Square attack. Compared to SimBA-DCT, PI-Attack with a schedule increased the ASR by 20.24\% and reduced the average query count by 100.9, while maintaining a similar average $\ell_2$ norm. For untargeted attacks, PI-Attack with a schedule improved the ASR by 8.74\% and 16.1\% over the Square attack and SimBA-DCT, respectively. A similar trend was observed with the PI-Attack on the medical Diabetic Retinopathy dataset.

\begin{figure}[h!]
\centering
\includegraphics[width=0.95\linewidth]{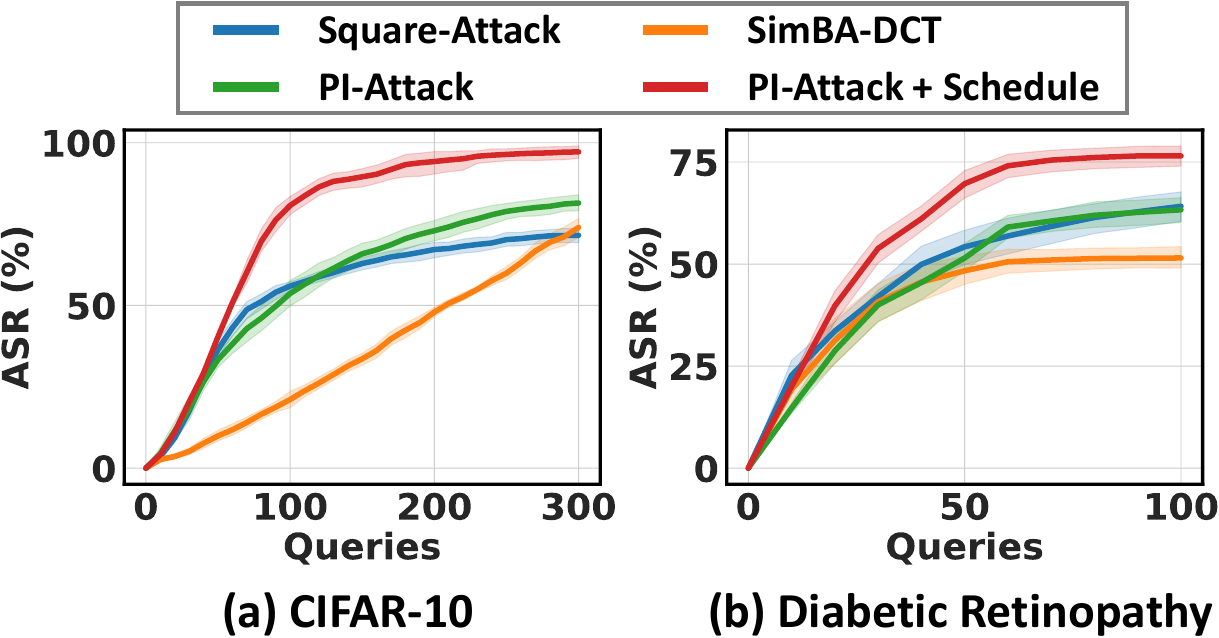}
\caption{Attack success rate (ASR) of different methods.}
\label{fig:attack}
\end{figure}

Figure~\ref{fig:attack} (a) and (b) illustrate the attack processes of our PI-Attack and previous methods on both the CIFAR-10 and Diabetic Retinopathy datasets. With a similar number of queries, our PI-Attack with scheduling consistently achieves a higher targeted ASR compared to SimBA-DCT and Square attack. This improvement can be attributed to the enhanced attack search efficiency realized through the perturbation scheduling and arc-shaped search order of the PI-Attack. Additionally, when the adversarial examples have the same $\ell_2$ norm, PI-Attack still manages to reach a higher ASR compared to other techniques. Specifically, on the CIFAR-10 dataset, PI-Attack with scheduling secures an ASR over 20\% higher than other methods while maintaining an average $\ell_2$ norm of approximately 3.0.

\noindent\textbf{RobPI Defense Evaluation.}
Table~\ref{t:results_defense} presents a comparison of the defense effects achieved by our RobPI and previous methods, including RND and RND-GF, as proposed by~\cite{qin2021random}. For targeted attacks on CIFAR-10, RND achieves a Defense Success Rate (DSR) of 39.22\% while maintaining a clean accuracy of 72.86\%. The RND-GF method, which includes Gaussian noise during training, reaches an defense success rate of 56.33\% with an accuracy of 73.71\%. In contrast to RND that incorporates noise into the input, our RobPI method introduces noise into the logits, which notably enhances the DSR by approximately 30\%. This improvement can be attributed to the fact that adding noise to the input of PNet with a polynomial activation function considerably diminishes the noise. However, introducing noise to the output bypasses this decay, as substantiated by our theoretical analysis of RobPI and empirical results. For example, RobPI-DNT registers a DSR of 91.88\% while maintaining a higher clean accuracy of 74.55\%. When compared to RND-GF in the context of targeted attacks on CIFAR-10, RobPI-DNT enhances DSR by 35.55\% and clean accuracy by 0.84\%.

Our RobPI and RobPI-DNT demonstrate consistent enhancements during untargted attacks and across other medical datasets. Specifically, RobPI-DNT shows a notable increase in untargeted defense success rate by 52.97\% compared to RND-GF on CIFAR-10. Likewise, on the Diabetic Retinopathy dataset, RobPI exhibits a rise of 14.54\% and 39.69\% in the targeted and untargeted defense success rates, respectively, over RND-GF, while achieving a 0.18\% increase in clean accuracy. The injection of noise into the input doesn't yield significant improvements in defense success rates. However, the DNT technique notably enhances both the defense success rate and clean accuracy.


\begin{figure}[h!]
\centering
\includegraphics[width=0.95\linewidth]{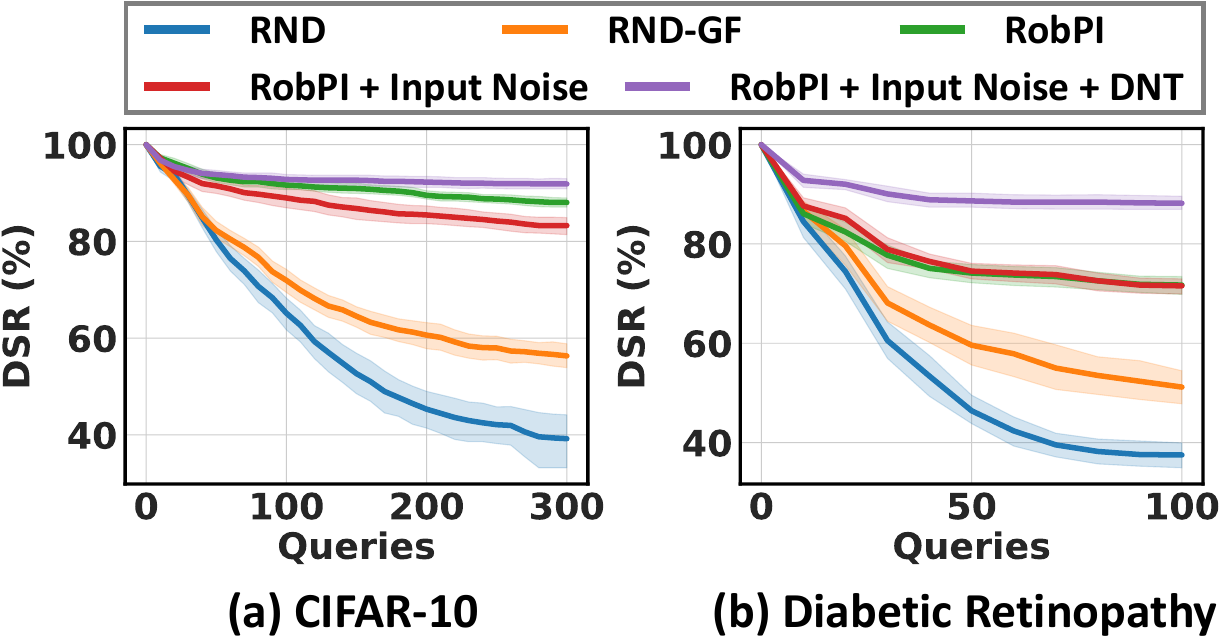}
 \caption{Defense success rate (DSR) of different methods.}
\label{fig:defense}
\end{figure}

Figure~\ref{fig:defense} presents the defense outcomes of prior works RND, RND-GF, and our techniques, including RobPI, RobPI+Input noise, and RobPI-DNT with input noise. Notably, all techniques display a high defense success rate in the initial queries due to the low attack success rate associated with a limited number of queries. However, as the number of queries increases, the defense success rates of both RND and RND-GF significantly decline. In contrast, our RobPI maintains a high defense success rate. Similar to RND-GF, the addition of noise to the input layer in RobPI+Input noise doesn't result in a significant improvement in defense. This suggests that the input noise might be diminished by the polynomial activation of PNet. With the incorporation of DNT techniques, RobPI-DNT further enhances the defense outcomes. It's important to note that without adding noise to the output layer, RobPI with input noise and DNT still fails to sustain a high defense effect.

\begin{figure}[h!]
\centering
\includegraphics[width=0.9\linewidth]{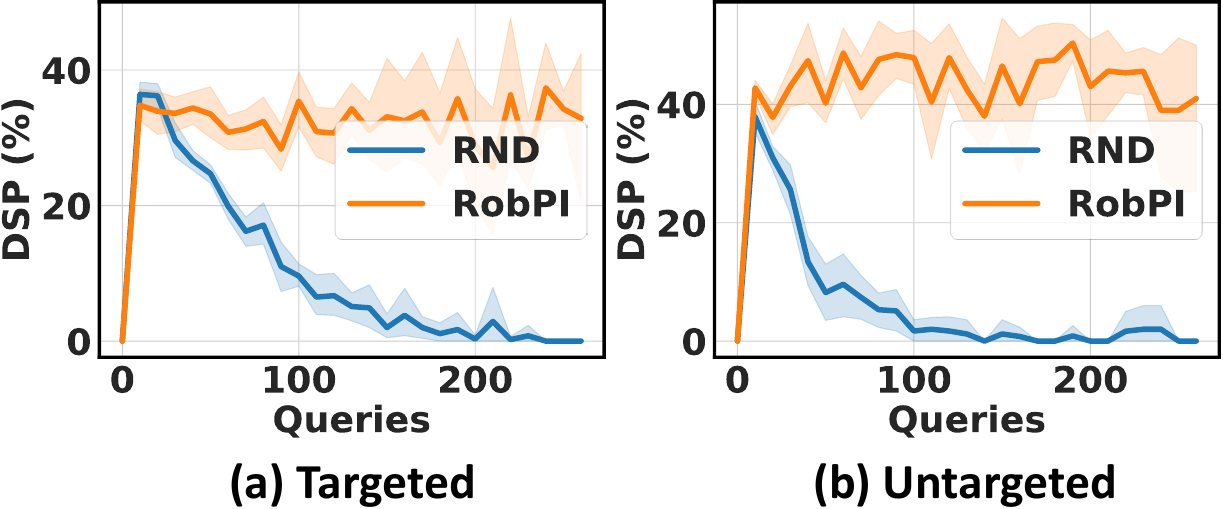}
 \caption{RobPI achieves a higher probability of disturbance success probability (DSP) than prior work RND.}
\label{fig:defense_performance}
\end{figure}
\begin{figure}[!ht]
  \centering
   \includegraphics[width=0.90\linewidth]{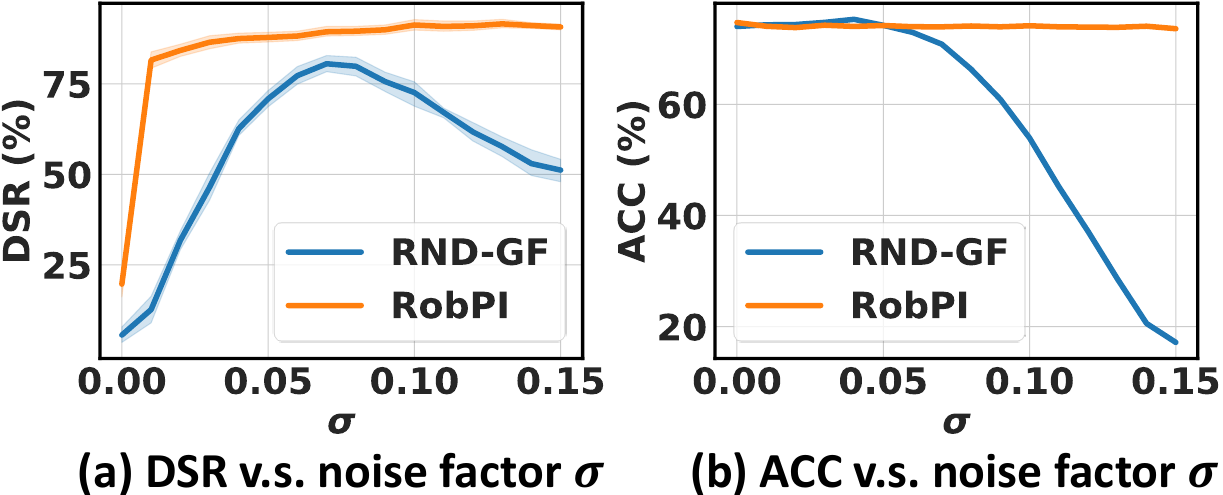}
   \caption{RobPI achieves a better balance between defense effect and clean accuracy than RND-GF.}
   \label{fig:disturb-sigma}
\end{figure}

Figure~\ref{fig:defense_performance} highlights the superior defense efficiency of RobPI in comparison to RND. By achieving a higher Disturbance Success Probability in both targeted and untargeted attacks, RobPI provides an empirical explanation for its heightened defense efficacy. Figure~\ref{fig:disturb-sigma} demonstrates that RobPI strikes a more effective balance between defense effect and accuracy than RND-GF. Specifically, for a given $\sigma$ (e.g., 0.1), Figure~\ref{fig:disturb-sigma}(a) shows that RobPI realizes a higher defense success rate than RND-GF. Similarly, Figure~\ref{fig:disturb-sigma}(b) indicates that, for the same $\sigma$, RobPI achieves a higher clean accuracy. RobPI therefore exhibits less noise sensitivity than RND-GF.

\noindent\textbf{Ablation Study.} As depicted in Figure~\ref{fig:disturb-sigma}, RobPI outperforms RND-GF in terms of striking a superior balance between defense proficiency and accuracy. Specifically, for a given $\sigma$ value (e.g., 0.1), Figure~\ref{fig:disturb-sigma}(a) reveals that RobPI yields a higher defense success rate compared to RND-GF. Meanwhile, Figure~\ref{fig:disturb-sigma}(b) indicates that, for an equivalent $\sigma$ value, RobPI secures higher clean accuracy. Consequently, RobPI demonstrates less sensitivity to noise than RND-GF.


\begin{table*}[t!]
\centering
\footnotesize
\setlength{\tabcolsep}{4pt}
\caption{The attack performance under different query numbers of Average PI-Attack.}
\begin{tabular}{ccccccccccccc}\toprule
\multirow{3}{*}{$N_q$} & \multicolumn{6}{c}{CIFAR-10} & \multicolumn{6}{c}{Diabetic Retinopathy} \\
\cmidrule(lr){2-7}\cmidrule(lr){8-13}
 & \multicolumn{2}{c}{Clean Accuracy} & \multicolumn{2}{c}{Average Queries} & \multicolumn{2}{c}{Defense Success Rate} & \multicolumn{2}{c}{Clean Accuracy} & \multicolumn{2}{c}{Average Queries} & \multicolumn{2}{c}{Defense Success Rate} \\
\cmidrule(lr){2-3}\cmidrule(lr){4-5}\cmidrule(lr){6-7}\cmidrule(lr){8-9}\cmidrule(lr){10-11}\cmidrule(lr){12-13}
 & Untar. & Target & Untar. & Target & Untar. & Target & Untar. & Target & Untar. & Target & Untar. & Target \\
\midrule
$1$ & \multicolumn{2}{c}{$74.55\%$} & $199.4$ & $299.7$ & $63.36\%$ & $91.88\%$ & \multicolumn{2}{c}{$68.09\%$} & $48.9$ & $52.0$ & $66.41\%$ & $88.67\%$ \\
$2$ & \multicolumn{2}{c}{$74.55\%$} & $200$ & $302$ & $62.19\%$ & $86.99\%$ & \multicolumn{2}{c}{$68.09\%$} & $49.5$ & $51.5$ & $\mathbf{58.53\%}$ & $\mathbf{71.85\%}$ \\
$3$ & \multicolumn{2}{c}{$74.55\%$} & $199.6$ & $299.3$ & $59.08\%$ & $\mathbf{83.52\%}$ & \multicolumn{2}{c}{$68.09\%$} & $48.3$ & $49.3$ & $60.88\%$ & $79.01\%$ \\
$4$ & \multicolumn{2}{c}{$74.55\%$} & $200$ & $300$ & $\mathbf{54.76\%}$ & $87.31\%$ & \multicolumn{2}{c}{$68.09\%$} & $50.5$ & $49$ & $69.09\%$ & $85.96\%$ \\
$5$ & \multicolumn{2}{c}{$74.55\%$} & $200.4$ & $299.8$ & $61.93\%$ & $91.98\%$ & \multicolumn{2}{c}{$68.09\%$} & $50.2$ & $48.4$ & $71.67\%$ & $89.44\%$ \\
\bottomrule
\end{tabular}
\label{t:attack_pro}
\end{table*}

\begin{table*}[t!]
\centering
\footnotesize
\setlength{\tabcolsep}{3pt}
\caption{The defense performance against Average PI-Attack using two proposed methods.}
\begin{tabular}{ccccccccccccc}\toprule
\multirow{3}{*}{Schemes} & \multicolumn{6}{c}{CIFAR-10} & \multicolumn{6}{c}{Diabetic Retinopathy} \\
\cmidrule(lr){2-7}\cmidrule(lr){8-13}
 & \multicolumn{2}{c}{Clean Accuracy} & \multicolumn{2}{c}{Average Queries} & \multicolumn{2}{c}{Defense Success Rate} & \multicolumn{2}{c}{Clean Accuracy} & \multicolumn{2}{c}{Average Queries} & \multicolumn{2}{c}{Defense Success Rate} \\
\cmidrule(lr){2-3}\cmidrule(lr){4-5}\cmidrule(lr){6-7}\cmidrule(lr){8-9}\cmidrule(lr){10-11}\cmidrule(lr){12-13}
 & Untar. & Target & Untar. & Target & Untar. & Target & Untar. & Target & Untar. & Target & Untar. & Target \\
\midrule
RobPI & \multicolumn{2}{c}{$74.55\%$} & $200$ & $299.3$ & $54.76\%$ & $83.52\%$ & \multicolumn{2}{c}{$68.09\%$} & $49.5$ & $51.5$ & $58.53\%$ & $71.85\%$ \\
$\mu=3$ & \multicolumn{2}{c}{$74.53\%$} & $200$ & $300$ & $64.19\%$ & $90.74\%$ & \multicolumn{2}{c}{$68.16\%$} & $50.5$ & $50.5$ & $64.96\%$ & $86.75\%$ \\
Penultimate layer & \multicolumn{2}{c}{$70.83\%$} & $201$ & $300.6$ & $59.92\%$ & $85.77\%$ & \multicolumn{2}{c}{$66.89\%$} & $50.5$ & $49$ & $61.56\%$ & $77.30\%$ \\
\bottomrule
\end{tabular}
\label{t:defense_pro}
\end{table*}

\section{Potential Attacks against RobPI}
\noindent\textbf{Average Inference Attacks.} 
In this section, we will examine a different attack: an adversary may send the same adversarial input multiple times and averages their outputs to bypass the RobPI's defense that adds Gaussian noise with zero means. We call this attack as \textit{average inference attack}. This type of attack is designed to exploit the fact that RobPI adds noise to the output probability distribution in each query, which has a zero mean and could potentially be averaged out by repeated queries.

Firstly, we investigate whether sending the same adversarial input multiple times and averaging their outputs can decrease the Defense Success Rate (DSR) of RobPI. By analyzing the results of this investigation in Table~\ref{t:attack_pro}, we can gain a better understanding of RobPI's robustness against this type of attack and identify potential countermeasures to further improve its defense effectiveness.

In our experiments, the number of queries for identical adversarial inputs ($N_q$) varied from 1 to 5. Tabel~\ref{t:attack_pro} reveals that for the CIFAR10 dataset, DSR decreases from $91.88\%$ to $86.99\%$ when $N_q$ is set to 2, and further decrease to $83.52\%$ when $N_q$ is set to 3. However, when $N_q$ is increased to 5, the DSR increases to $91.98\%$.

This trend can be explained by the fact that when $N_q$ is increased, the attacker needs to query $N_q$ times to add noise once, which means that the total number of queries available for adding noise decreases as $N_q$ increases. Therefore, with a fixed number of total queries, the attacker will add less noise when $N_q$ is larger. For example, under a fixed total number of queries, e.g., 300, an attacker can add noise up to 300 times when $N_q=1$ but only up to 60 times when $N_q=5$.

\noindent\textbf{Defense of Average Inference Attacks.} 
In this section, we propose two defense methods to counteract the average inference attack. The first method suggests using a Gaussian distribution with a non-zero mean ($\mu$) instead of a standard Gaussian distribution. The second method recommends adding noise at the penultimate layer rather than at the last layer. To assess their effectiveness, we measured RobPI's DSR when ($\mu$) was set to 3 and when noise was added at the penultimate layer. We determined that $N_q$ should be set to values that result in lower DSRs based on Table 3: $N_q=4$ for untargeted attacks against CIFAR10, $N_q=3$ for targeted attacks on CIFAR10, and $N_q=2$ for both untargeted and targeted attacks on Diabetic Retinopathy.

Based on the results presented in Table~\ref{t:defense_pro}, we can observe that using a non-zero mean Gaussian distribution renders the Average PI-Attack ineffective while maintaining a high Clean Accuracy (ACC) similar to that of using a zero mean Gaussian distribution. For instance, on the CIFAR10 dataset, the DSR of targeted attacks increased from $83.52\%$ to $90.47\%$, and the DSR of untargeted attacks increased from $54.76\%$ to $64.19\%$, while the ACC remained at $74.53\%$. Similarly, adding noise at the penultimate layer instead of the last layer also improved DSR. For example, on CIFAR10, the DSR of targeted attacks increased from $83.52\%$ to $85.77\%$, and the DSR of untargeted attacks increased from $54.76\%$ to $59.92\%$.

\section{Conclusion}

This work first introduces PI-Attack, an innovative inference manipulation attack for private inference protocols reliant on fully homomorphic encryption. The attack necessitates $3\times \sim 8\times$ fewer queries than current approaches. Moreover, we present RobPI, a robust private inference protocol designed to withstand malicious clients, achieved by incorporating cryptography-compatible noise in the feature and logits layers and deploying a DNT technique. RobPI demands over $10\times$ more query numbers compared to previous defense methods, substantially elevating the attack difficulty. Theoretical analysis and empirical testing show that RobPI can diminish the attack success rate by approximately $91.9\%$ across various neural networks and datasets. 

\section*{LLM Usage Considerations}
\addcontentsline{toc}{section}{LLM Usage Considerations}

LLMs were used for \emph{editorial purposes only} in this manuscript, specifically to check grammar, spelling, and minor typographical issues. All generated outputs were carefully inspected by the authors to ensure accuracy, originality, and compliance with the scientific content. No part of the methodology, experiments, or research contributions relied on LLMs.

\appendix

\subsection{DCT and IDCT}

  

The discrete cosine transform (DCT) is a mathematical tool that can be employed to shift an image from the spatial domain to the frequency domain. By initiating a search from lower frequencies and progressing to higher ones, one can effectively pinpoint an adversarial sample, thereby reducing the number of required queries. DCT represents an image as a sum of sinusoids of varying magnitudes and frequencies. Specifically, for an input image $X \in \mathbb{R}^{d\times d}$, the DCT transform $V=DCT(X)$ is:
\begin{equation} \small
    V_{m,n}=\alpha_m\alpha_n\sum_{i=0}^{d-1}\sum_{j=0}^{d-1}X_{i,j}\cos{\frac{\pi(2i+1)m}{2d}}\cos{\frac{\pi(2j+1)n}{2d}}
    \label{eq:schedule_1}
\end{equation}
where
\begin{equation} \small
    \alpha_m= \begin{cases}
    \sqrt{\frac{1}{d}},\quad &m=0 \\
    \sqrt{\frac{2}{d}},\quad &1\leq m\leq d-1
    \end{cases} 
    \label{eq:schedule_2}
\end{equation}
and
\begin{equation} \small
    \alpha_n= \begin{cases}
    \sqrt{\frac{1}{d}},\quad &n=0 \\
    \sqrt{\frac{2}{d}},\quad &1\leq n\leq d-1
    \end{cases} 
    \label{eq:schedule_3}
\end{equation}
for $0 \leq m,n \leq d-1$.

The values $V_{m,n}$ are called the DCT coefficients of. The DCT is an invertible transform, and its inverse IDCT is given by:
\begin{equation} \small
    X_{i,j}=\sum_{m=0}^{d-1}\sum_{n=0}^{d-1}\alpha_m\alpha_nV_{m,n}\cos{\frac{\pi(2i+1)m}{2d}}\cos{\frac{\pi(2j+1)n}{2d}}
    \label{eq:schedule_4}
\end{equation}
for $0 \leq i,j \leq d-1$. The basis functions are:
\begin{equation} \small
    \alpha_m\alpha_n\cos{\frac{\pi(2i+1)m}{2d}}\cos{\frac{\pi(2j+1)n}{2d}}
    \label{eq:schedule_5}
\end{equation}

\begin{figure}[h!]
  \centering
   \includegraphics[width=0.5\linewidth]{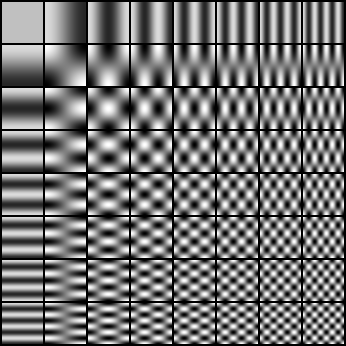}
   \caption{The 64 Basis Functions of an 8-by-8 Image.
   }
   \label{fig:DCT and IDCT}
\end{figure}

The inverse DCT (IDCT) is an inverse phase of DCT. The IDCT equation can be interpreted as meaning that any $d\times d$ image can be written as a sum of basis functions. The DCT coefficients $V_{m,n}$ can be regarded as the weights applied to each basis function, with lower frequencies represented by lower $m,n$. Especially for $8\times 8$ images, the 64 basis functions are illustrated by Figure~\ref{fig:DCT and IDCT}.
Horizontal frequencies increase from left to right, and vertical frequencies increase from top to bottom.

\subsection{Cryptosystems Settings}
For cryptosystems of RobPI, one can follow the LoLa~\cite{Brutzkus:ICML19}, where the BFV scheme in SEAL~\cite{sealcrypto} is used. For MNIST and CIFAR-10, the plaintext modulus $m=2148728833\times 2148794369\times 2149810177$, modulus degree $N=16384$, coefficient modulus $q=\sim440$ bits. The security level is larger than 128 bits which are verified by $lwe\_estimator$~\cite{lwe_estimator}. To run RobPIs, one can run all experiments on the same Azure standard B8ms virtual machine with 8 vCPUs and 32GB DRAM.

\subsection{Results on MNIST dataset}
\begin{figure}[h!]
  \centering
   \includegraphics[width=\linewidth]{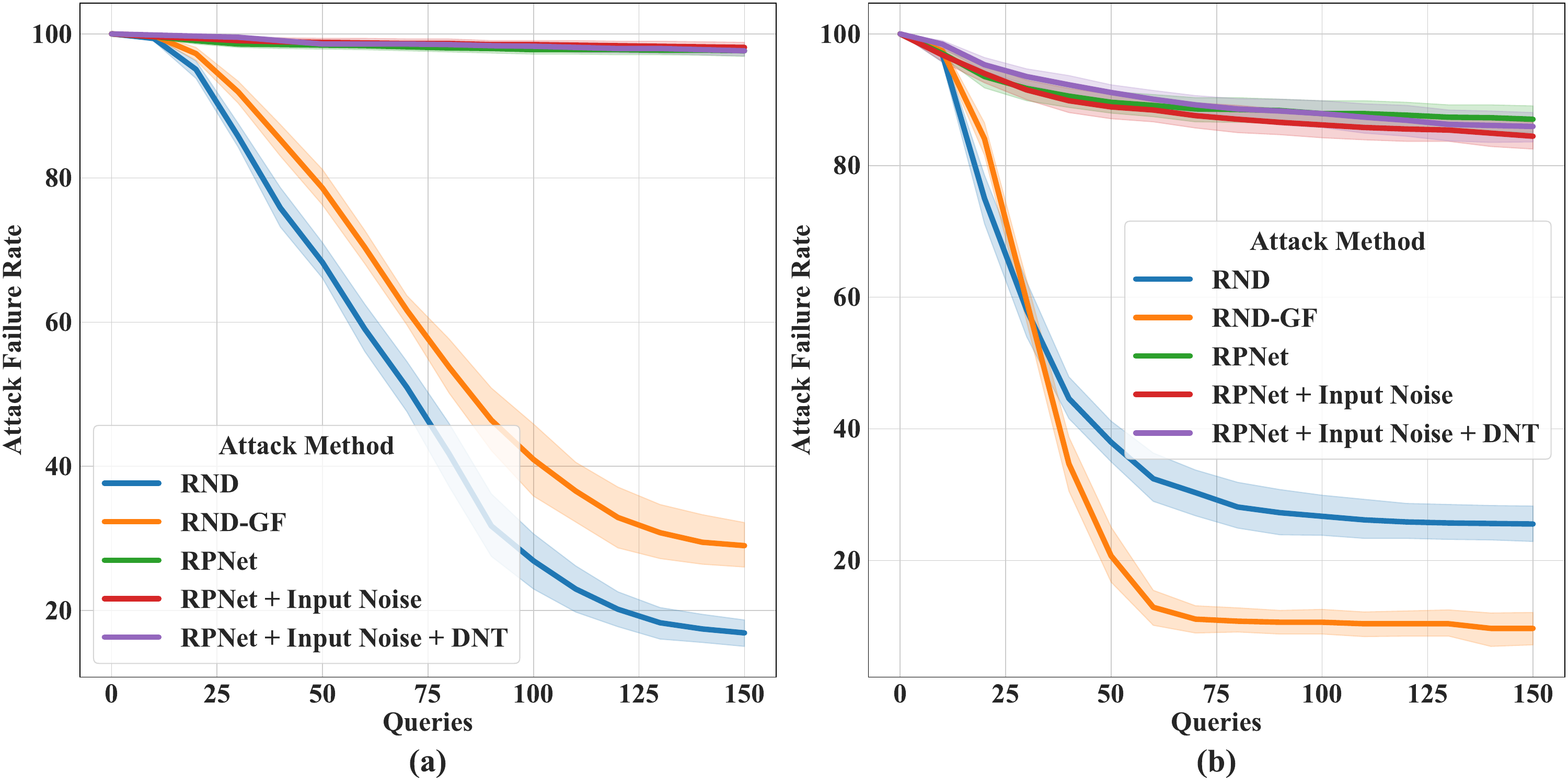}
   \caption{defense results on MNIST.
   }
   \label{fig:MNIST defense}
\end{figure}
In Figure~\ref{fig:MNIST defense}, we compare different defense methods on MNIST. We show that compared with the traditional methods, the DSR of RobPI proposed in our paper is greatly improved.


\balance
\bibliography{example_paper}
\bibliographystyle{IEEEtran}
\end{document}